\documentclass{ws-procs9x6-cpt16}
\begin{document}

\newcommand{\refeq}[1]{(\ref{#1})}
\def\etal {{\it et al.}}

\title{Emergence of Gauge Invariance from Nambu Models}

\author{L.F.\ Urrutia }

\address{Instituto de Ciencias Nucleares,
Universidad Nacional Aut\'onoma de M\'exico\\
Apartado Postal 70-543, Delegaci\'on Coyoac\'an,
Ciudad de M\'exico, 04510, M\'exico}

\begin{abstract}
In the framework of a hamiltonian nonperturbative approach we show that
after demanding current conservation together with the Gauss constraints
at some initial time in a nonabelian Nambu model, we recover
the corresponding Yang-Mills theory. In this way,
the spontaneous Lorentz symmetry breaking present in the Nambu model becomes
unobservable and the Goldstone modes
can be identified with the corresponding gauge bosons.
\end{abstract}

\bodymatter

\section{Introduction}
\label{aba:sec1}

Up to now the search for Lorentz invariance violation (LIV) has been very
elusive, in spite of the tremendous efforts for increasing the precision in
the experimental setups. In this contribution
we take the opposite approach, in which spontaneous Lorentz symmetry
breaking (SLSB) arising from a more fundamental theory does
not lead to observable LIV, but rather provides a dynamical interpretation
of the corresponding massless gauge bosons in terms of the Goldstone modes (GM)
arising from such spontaneous breaking. This is an old idea pioneered by
Dirac,\cite{PAMD} Bjorken,\cite{JDB} and Nambu, \cite{YN} which has a long history in the
literature.\cite{others} We will not pay attention to the mechanism which induces the
SLSB and we will just focus on the low energy approximation, where only
the GM are excited. The most economic way to do this is by
starting from so called nonabelian Nambu models (NANM), which constitute generalizations of the original Nambu model.\cite{YN} They are defined
by a Yang-Mills lagrangian, to be called the mother gauge theory (MGT), plus
a nonlinear constraint (NLC) encoding the SLSB, which should be solved and substituted into the lagrangian. This formulation is analogous to the
description of pions in terms of a nonlinear sigma model. Let us
start from the following NANM
\begin{equation}
\mathcal{L}=-\frac{1}{4}F_{\mu \nu }^{a}F^{a\mu \nu }-J^{a\mu }A_{\mu
}^{a},\qquad A^{a\mu }A_{\mu }^{a}=n^{2}M^{2},\quad D_{\mu }J^{a\mu }=0,
\label{aba:EQ1}
\end{equation}
coupled to a conserved current $J^{a\mu }$. Here $a=1,\dots,N$ are $SU(N)$ labels, $M$ is the SLSB scale and
the constant group vector $n_{\mu }^{a}$ indicates the direction of the
vacuum along which the SLSB occurs. In this way, one interprets the
nonlinear constraint as describing a nonzero vacuum expectation value $
\langle \hat{A}_{\mu }^{a}\rangle =n_{\mu }^{a}M$. The $(3N-1)$ equations of motion arising from the lagrangian \refeq{aba:EQ1} can be presented in the form
\begin{equation}
{\cal E}^{a 0}+{\cal E}^{1 3}\frac{A_0^a}{A_1^3}=0, \quad
{\cal E}^{a i}-{\cal E}^{1 3}\frac{A_i^a}{A_1^3}=0, \quad i\neq 3, a\neq 1,
\label{aba:EQ0}
\end{equation}
with the notation ${\cal E}^{a \nu}=(D_\mu F^{\mu\nu}-J^\nu)^a$. Here $F^a_{\mu\nu}$ is the nonabelian Faraday tensor and $D_\mu$ denotes the covariant derivative. Let us observe that the $4N$ equations of motion for the MGT are ${\cal E}^{a \nu}=0$, which can be obtained from Eqs.\ \refeq{aba:EQ0} by demanding ${\cal E}^{a 0}=0$, for example.
Let us emphasize that we have two theories with different equations of motion and properties: the NANM is a theory with no gauge invariance having $3N$ d.o.f., coupled to a conserved current, while the MGT is a standard gauge theory with $N$ first class constraints, $2N$ d.o.f. yielding a conserved current.\cite{EU}

The purpose of this note is to determine and clarify the conditions under which a NANM turns out to be equivalent to its MGT.

Our main motivation arises from the perturbative calculations previously performed in the abelian case, where a highly nonlinear theory results after the constraint and the lagrangian are written in terms of the field redefinition $A_\mu=n_\mu+ a_\mu, \,\, n\cdot a=0$,
with $a_\mu$ describing the GM excitations. In the tree and one-loop approximations, the result for the amplitudes for some specific processes is that all LIV contributions cancel yielding the standard results in QED.\cite{YN,AZ}
\section{The non-perturbative hamiltonian approach}
\label{aba:sec2}
Following Ref.\ \refcite{EU}, the strategy to find the conditions for equivalence is the following: (1) We take as a benchmark the well-known canonical algebra and hamiltonian of the MGT, in terms of the basic canonical variables $A_i^a, E^{bj}, j=1,2,3$.
(2) After solving the NLC we determine the canonical algebra and hamiltonian of the NANM. The most direct (and cumbersome) way to proceed is to solve the NLC for one variable, $A^1_3$ for example, and start the construction from the remaining $(4N-1)$ coordinates, with the presence of $2(N-1)$ second class constraints. Here we present a convenient parametrization of the NLC that drastically simplifies the procedure. (3) In any case, it is always possible to express the canonical variables of the NANM in terms of those of the MGT. In this way, we are able to rewrite the canonical algebra and hamiltonian of the NANM in terms of the canonical variables of the MGT. (4) The comparison of this transcription with the known results in the MGT allows us to establish the required conditions for equivalence. Let us introduce the $3N$ coordinates $\Phi^a_i$, where we split the indices $i=1,2,3$ into ${\bar i}=1,2$ and $3$, together with the parametrization\cite{EU}
\begin{equation}
A_0^a=\Phi^a_3 \left(1+\frac{N}{4\, \Phi^a_3 \Phi^a_3}\right), \,\, A_3^a=\Phi^a_3 \left(1-\frac{N}{4\, \Phi^a_3\Phi^a_3}\right),
\label{aba:EQ2}
\end{equation}
with $N=\Phi^a_{\bar k}\Phi^a_{\bar k}+n^2 M^2$. In terms of the new coordinates the NLC in Eq.~\refeq{aba:EQ1} is identically satisfied. Let us focus now on the invertible change of variables $A^a_i=A^a_i(\Phi_j^b)$, which yields
${\dot A}^a_i={\dot A}^a_i(\Phi, {\dot \Phi})$ and ${\dot A}^a_0={\dot A}^a_0(\Phi, {\dot \Phi})$, recalling that we also know $A^a_0=A^a_0(\Phi)$. We observe that we will not require the explicit form of the transformations in the following. In this way, the lagrangian of the NANM can be written as
${\cal L}_{\rm NANM}(\Phi, {\dot \Phi})$. The coordinates $\Phi^a_i$ together with the NANM canonically conjugated momenta $\Pi^{b j}$ satisfy the standard canonical algebra and define a regular system (i.e., no constraints appear in this parametrization). The usual definition of the nonabelian electric field, $E^a_i=F^a_{0i}={\dot A}^a_i-D_iA^a_0$, allows us to obtain the relations
\begin{equation}
\Pi^b_i=\frac{\partial A^a_j}{\partial \Phi^b_i} E^a_j, \qquad
E^b_i=\frac{\partial \Phi^a_j}{\partial A^b_i} \Pi^a_j,\qquad \Pi^a_i {\dot \Phi}^a_i=E_j^b{\dot A}^b_j.
\label{aba:EQ3}
\end{equation}
Next we calculate the NANM hamiltonian ${\cal H}_{\rm NANM}(\Pi,\Phi)=\Pi{\dot \Phi}-{\cal L}_{\rm NANM}(\Phi, {\dot \Phi})$ and explicitly verify that it can be written as
\begin{equation}
{\cal H}_{\rm NANM}=\frac{1}{2}\left(E^a_iE^a_i+B^a_iB^a_i \right)+J^{a i}A^a_i
-\left( D_i E^b_i-J^{0 b}\right)A^b_0,
\label{aba:EQ4}
\end{equation}
with $B^a_i=\epsilon_{ijk}F^a_{jk}/2$. Here $E^a_i, B^a_i, A^a_\mu$ are just labels for the corresponding functions of $(\Phi, \Pi)$, in terms of which it will be simpler to make contact with the MGT. Also, these relations would allow the explicit calculation of algebra among $A^a_i, E^{b i}$, in terms of the NANM canonical algebra. Let us observe that in the hamiltonian form of the action for the NANM, the last relation in Eq.\ \refeq{aba:EQ3} identifies $E^{a i}$ as the canonically conjugated momenta to the coordinates $A^a_i$ of the MGT. Moreover, since the transformation among the variables $(E, A)$ and $(\Pi, \Phi)$ is generated only by the coordinate transformations $A=A(\Phi)$, we can assert that this is indeed a canonical transformation.\cite{EU} In this way we recover the canonical algebra for the MGT phase-space variables $A^a_i, E^{b i}$ without any further calculation. Summarizing, up to now we have proved that the NANM canonical algebra yields the canonical algebra of the MGT. Nevertheless, both theories are still not equivalent because Eq.\ \refeq{aba:EQ4} is not the hamiltonian of the MGT, in spite of its almost identical form. In fact, $A_0^b$ are not an arbitrary functions, but only labels for some combinations of the coordinates in the NANM. Also, the Gauss functions $G^b \equiv \left( D_i E^b_i-J^{b 0}\right)={\cal E}^{b 0}$ are not constraints, as they should be in a gauge theory.
To proceed, we calculate the time evolution of the Gauss functions under the NANM dynamics, obtaining
\begin{equation}
{\dot G}^b= -D_\mu J^\mu + M^{bc}G^c + D_i(N^{ibc}G^c),
\label{aba:EQ5}
\end{equation}
where $M^{bc}, N^{ibc}$ are known functions of NANM phase space. To recover gauge invariance it is enough to impose the Gauss constraints $G^b=0$, together with current conservation $D_\mu J^\mu=0$ at a given time slice $t_0$. Then Eq.\ \refeq{aba:EQ5} guarantees that $G^b=0$ at $t_0+ \delta t$. Then, according to Eqs.~\refeq{aba:EQ0}, the equations of the MGT ${\cal E}^{a \mu}=0$ must be valid at $t_0+ \delta t$. Using the time independent identity $(D_\mu D_\nu F^{\mu\nu})^a=0$ we then obtain $D_\mu J^\mu=0$ for $t_0+ \delta t$. Thus, iterating this procedure from the initial conditions we obtain current conservation and the Gauss constraints for all time. The latter can be added to the hamiltonian \refeq{aba:EQ4} as additional constraints via arbitrary functions $R^b$ which now replace $A_0^b$ by $S^b= A_0^b + R^b$. The canonical algebra, together with current conservation, guarantees that no additional constraints appear. Summarizing, under the initial conditions already specified, the equivalence between the NANM and the MGT is established.

\section*{Acknowledgments}
L.F.U. is partially supported by the project CONACyT $\#$ 237503 and the project DGAPA-UNAM $\#$ IN-104815.


\begin{thebibliography}{x}

\bibitem{PAMD}
P.A.M.\ Dirac,
Nature (London) {\bf 168}, 906 (1951).

\bibitem{JDB}
J.D.\ Bjorken,
Ann.\ Phys.\ (N.Y.) {\bf 24}, 174 (1963).

\bibitem{YN} Y.\ Nambu,
Prog.\ Theor.\ Phys.\ Supp.\ {\bf E68}, 190 (1968).

\bibitem{others}
P.\ Krauss and E.T.\ Tomboulis,
Phys.\ Rev.\ D {\bf 72}, 045015 (2002);
R.\ Righi and G.\ Venturi,
Int.\ J.\ Theor.\ Phys.\ {\bf 21}, 63 (1982);
J.L.\ Chekareuli, C.D.\ Froggatt and H.B.\ Nielsen,
Nucl.\ Phys.\ B {\bf 821}, 65 (2009);
R.\ Bluhm, N.L.\ Gagne, R.\ Potting and A.\ Vurblevskis,
Phys.\ Rev.\ D {\bf 77}, 125007 (2008);
V.A.\ Kosteleck\'y and R.\ Potting,
Phys.\ Rev.\ D {\bf 79}, 065018 (2009);
S.M.\ Carroll, H.\ Tam and I.K.\ Wehus,
Phys.\ Rev.\ D {\bf 80}, 0250290 (2009);
C.A.\ Escobar and L.F.\ Urrutia,
Phys.\ Rev.\ D {\bf 92}, 025042 (2015).

\bibitem{EU}
C.A.\ Escobar and L.F.\ Urrutia,
Phys.\ Rev.\ D {\bf 92}, 025013 (2015).

\bibitem{AZ}
A.T.\ Azatov and J.L.\ Chkareuli,
Phys.\ Rev.\ D {\bf 73}, 065026 (2006).

\end{thebibliography}
\end{document}